\documentclass{jkas}
\usepackage{amsmath}

\def\beginpage{1} 
\def\received{February 30, 2014} 
\def\accepted{February 31, 2014} 
\date{Received \received ; accepted \accepted}

\title{
NVST Data Archiving System Based On FastBit NoSQL Database
}

\author[1,2,3]{Feng~Wang}
\author[2,3]{Ying-bo~Liu}
\author[1]{Kai-fan~Ji}
\author[1]{Hui~Deng}
\author[1,2,3]{Wei~Dai}
\author[1]{Bo~Liang}

\affil[1]{Computer Technology Application Key Lab of Yunnan Province, Kunming University of Science and Technology, Chenggong, Kunming, 650500, China; \email{wangfeng@acm.org}}
\affil[2]{Yunnan Observatories, Chinese Academy of Sciences, Kunming 650011, China;}
\affil[3]{University of Chinese Academy of Sciences, Beijing, 100049, China; \email{liuyingbo@ynao.ac.cn}}

\def\corrauthor{
Feng~Wang
}

\def\runningauthor{
Ying-bo Liu ET AL.
}

\def\runningtitle{
The NVST Data Archiving System
}

\def\keywords{
Fastbit database --- index  --- query
}

\def\abstracttext{

The New Vacuum Solar Telescope (NVST) is a 1-meter vacuum solar telescope that aims to observe the fine structures of active regions on the Sun. The main tasks of the NVST are high resolution imaging and spectral observations, including the measurements of the solar magnetic field. The NVST has been collecting more than 20 million FITS files since it began routine observations in 2012 and produces a maximum observational records of 120 thousand files in a day. Given the large amount of files, the effective archiving and retrieval of files becomes a critical and urgent problem. In this study, we implement a new data archiving system for the NVST based on the Fastbit NoSQL database. Comparing to the relational database (i.e., MySQL), the Fastbit database manifests distinctive advantages on indexing and querying performance. In a large scale database of 40 million records, the multi-field combined query response time of Fastbit database is about 15 times faster and fully meets the requirements of the NVST. Our study brings a new idea for massive astronomical data archiving and would contribute to the design of data management systems for other astronomical telescopes.
}

\begin{document}
\jkashead 


\section{INTRODUCTION}
\label{sec:relate}
The New Vacuum Solar Telescope (NVST) is a 1-meter vacuum solar telescope which is located at the northeast side of Fuxian Lake, a world-class observational site in Yunnan province of China. The main tasks of the NVST are high resolution imaging and spectral observations in visible and near infrared bands, including measurements of the solar magnetic field \citep{Liu20111meter,liu2014new}.

As a modern optical telescope, NVST has been producing massive data of 20 million observation files since it began routine observation in 2012. The average number of records in one year is about 10 to 12 millions. Such massive data are produced by a variety of instruments on the NVST. The NVST has a multi-channel high resolution imaging system, one chromosphere channel and two photosphere channels. The fine structures of active regions and their evolutions in the photosphere and the chromosphere can be observed at the same time. The NVST also has two multi-band spectrometers, a medium-dispersion multi-band spectrometer in optical bands and a high-dispersion multi-band spectrometer in the near infrared.

\begin{figure}[!hb]
\centering
\includegraphics[width=80mm]{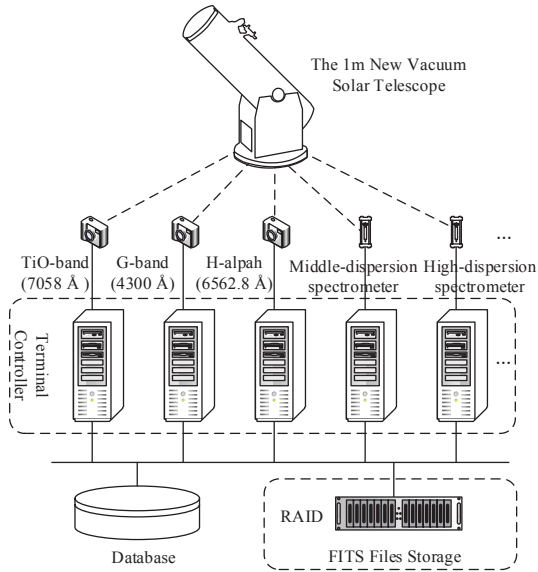}
\caption{The storage diagram of observational data files.\label{fig-struct}}
\end{figure}

Figure~\ref{fig-struct} shows the architecture diagram of the NVST data storage. The observational data produced by the various instruments are hierarchically saved in FITS file format into a massive storage system. To avoid conflicts, multi-level directories are setup to store the respective files. The top level is a directory named using the observation date. The second level consists of the subdirectories under the top level directory named after the different terminal device ID. The FITS files are saved into different subdirectories depending on the ID of the terminal device.

It is necessary to build a data archiving system to manage such an enormous amount of observation files. Obviously, it is very difficult for astronomers to directly query the data from the massive FITS files. There is therefore an urgent need for a the data archiving system(DAS) that provides the functions of data storage, indexing and retrieval in a simple and intuitive manner. NVST-DAS has its specific requirements in building the DAS, such as three-year data online storage and high performance data retrieval. This means the amount of the online data records would reach at least 40 millions. High-performance data retrieval would benefit the process of high-resolution image reconstruction, a typical use of the NVST-DAS by astronomers is to quickly sort the possible high quality images. Considering that the number of the NVST observational data is increasing so fast, we are also preoccupied with questions of performance of data retrieval, especially in terms of the scalability of the database.

The database technology is the fundamental platform of the DAS. Among modern database technologies, the relational database technology is mature and reliable. Lots of open source data archiving systems would help astronomers to build a data archiving system easily and quickly. Although the relational database technology is treated as the first choice to build this kind of system and has been discussed extensively in the literature, several studies \citep{otoo2010accelerating,parker2006comparing,Jatana2012survey,leavitt2010will} also regard that the relational database technique has limitations for large scale data indexing and retrieval. The continued increase of the amount of records might result in a noticeable performance penalty in data querying. It has become a very challenging problem to build an efficient management system for large astronomical databases with relational database technology. Although many strategies could be considered (e.g., splitting into sub-databases), it would result in a great deal of maintenance work and many potential consistency problems in the observational data. Apart from the relational database technology, NoSQL (Not Only SQL) database technology is an another choice of building the DAS. However, there are few successful cases in astronomical data archiving.

In this study, we focus on the possibility of practical applications of the NoSQL database technology for massive astronomical archiving. We investigate the Fastbit NoSQL database technology theoretically, and experimentally and compare it to the relational database technology. The results of our contrast experiments prove that the Fastbit database provides significantly higher performance of data retrieval from large amount of records than a traditional relational database.

The paper is organized as follows: After an introduction of research background in Section 1, we make a survey of previous works in the literature concerned with massive data archiving in Section 2. In Section 3, we make a theoretical analysis on the bitmap index technique and further conduct an availability judgment of bitmap index through a series of contrast experiments. The NVST-DAS is introduced in more detail in Section 4 describing the architecture design of the DAS and the related key modules. Our conclusions are offered in the last section.

\section{RELATED WORKS}
\label{sec:relateworks}

The archiving of observational data is one of the oldest and most fundamental issues in astronomical observation. In general, the archiving system is typically a complex management information system, on which both the data indexing and retrieval performance heavily depend on the data indexing and retrieval performance. With the advent of computer database technology, many emerging database technologies were for the first time used for astronomical data archiving.

Based on the difference of indexing and retrieval techniques, we investigate in detail and classify previous studies involving astronomical data indexing, retrieval and archiving into two categories:  Structured Query Language (SQL) based and Not Only Structured Query Language (NoSQL) based.

{\em1. SQL based.} Due to the advantages of the relational database technology, many astronomical data archiving systems are have been created using this technology. Open source relational database products such as MySQL and PostgreSQL are widely used in astronomical data processing. For example, Atmospheric Imaging Assembly (AIA) is one of three instruments aboard the Solar Dynamics Observatory (SDO) \citep{pesnell2012the} and takes 4096$\times$4096-pixel images every 3/4 of a second\footnote{http://spacemath.gsfc.nasa.gov/weekly/6Page137.pdf}. Therefore, the number of the AIA images in one day is about 115 thousand which is the same as the number of NVST images. To retrieve the data, a data record management system was developed with the PostgreSQL database\footnote{http://www2.mps.mpg.de/projects/seismo/GDC-SDO/}.

Generally, the relation model of the observational data should be firstly considered \citep{baba2002development, szalay2000designing,cui2008search}. All fields and tables should be designed carefully in order to normalize all tables in the database to a 3rd Normal Form. The field of observational date and time is typically defined as the primary key so as to identify a specific observation record. The structured query language is used to retrieve the data from the database.

{\em2. NoSQL based.} As ever increasing volumes of data (which some call the digital landfill) are generated, it becomes nearly impossible to store and retrieve data with a single computer. The management of large volumes of astronomical FITS data files is beyond the scope of traditional data indexing technologies \citep{greisen1981extension, gray2005scientific, chou2011parallel, parkerwood2013examining}. Thus, NoSQL technology was presented, offering a timely alternative to the object-relational data model.

NoSQL commonly uses a hash index to deal with massive data querying \citep{strauch2011nosql}. Spatial indexing, such as HTM and HEALPix, is successfully used in astronomical object's indexing and large multi-dimensional databases \citep{kunszt2000indexing,gorski2005healpix}. \citet{baruffolo1999r} implemented prototypes of R-Tree and B-Tree for astronomical data indexing. \citet{fu2012Indexing} studied a novel approach to index astronomical objects.

A bitmap index is a special database index proposed for the first time for the commercial database Model 204 by \citet{neil1989model}. The following studies on bitmap index showed that bitmap indexing is well suited for data querying of large scientific data \citep{otoo2010accelerating}. \citet{sinha2006bitmap} proposed an adaptive, multi-level and multi-resolution bitmap indexing scheme for scientific data index.

However, the bitmap index is known to be efficient for attributes each with few unique values(low-cardinality). \citet{wu2001notes} invented the compressed word-aligned hybrid bitmap (WAH) index scheme which has been proven to be more efficient in extremely large data both in theory and in practice. Especially WAH compressed bitmap indexing is efficient for high cardinality attributes \citep{ wu2005FastBit,wu2006optimizing,wu2008breaking}. Based on these preliminary studies, Fastbit \citep{wu2005FastBit} was developed and released as a typical bitmap indexing software development library built on an optimized compressed word-aligned algorithm. FastQuery \citep{chou2011fastquery} is a high-level data query engine based on the Fastbit library and has been implemented successfully in scientific domains such as particle accelerator and global climate.

Only a few studies discussed the applications of Fastbit/FastQuery in astronomy \citep{ma2012efficient}. The reason is that although FastQuery provides an interface for the FITS file format, the interface has never been really implemented \citep{chou2011fastquery}. FastQuery mainly focus on the high performance indexing and retrieval of other scientific datafile formats, e.g., the HDF5 \citep{folk2011overview}. It has a similar structure to the FITS format but no mechanism to support data indexing directly \citep{gosink2006hdf5}.

In summary, our investigation shows that there are numerous previous works in the literature that have studied the scientific data indexing and high performance retrieval. However, only a few of them discussed the high performance data retrieval approach under the condition of massive data. Although Fastbit NoSQL database has been integrated and deployed in many scientific research fields, there are only a few studies for its implementation in astronomical data processing and few works in astronomical data achieving. It is necessary to have a practical attempt on using bitmap index for high performance astronomical data indexing and retrieval.

\section{THEORETICAL ANALYSES AND EXPERIMENTS}
\subsection{Bitmap Index}
A bitmap index is a special kind of database index that uses bitmaps. Bitmap indexes use bit arrays (commonly called bitmaps) and answer queries by performing bitwise logical operations on these bitmaps. Bitmap indexes have a significant space and performance advantage over other structures for querying of low-cardinality data.

\begin{figure}[!htb]
\centering
\includegraphics[width=80mm]{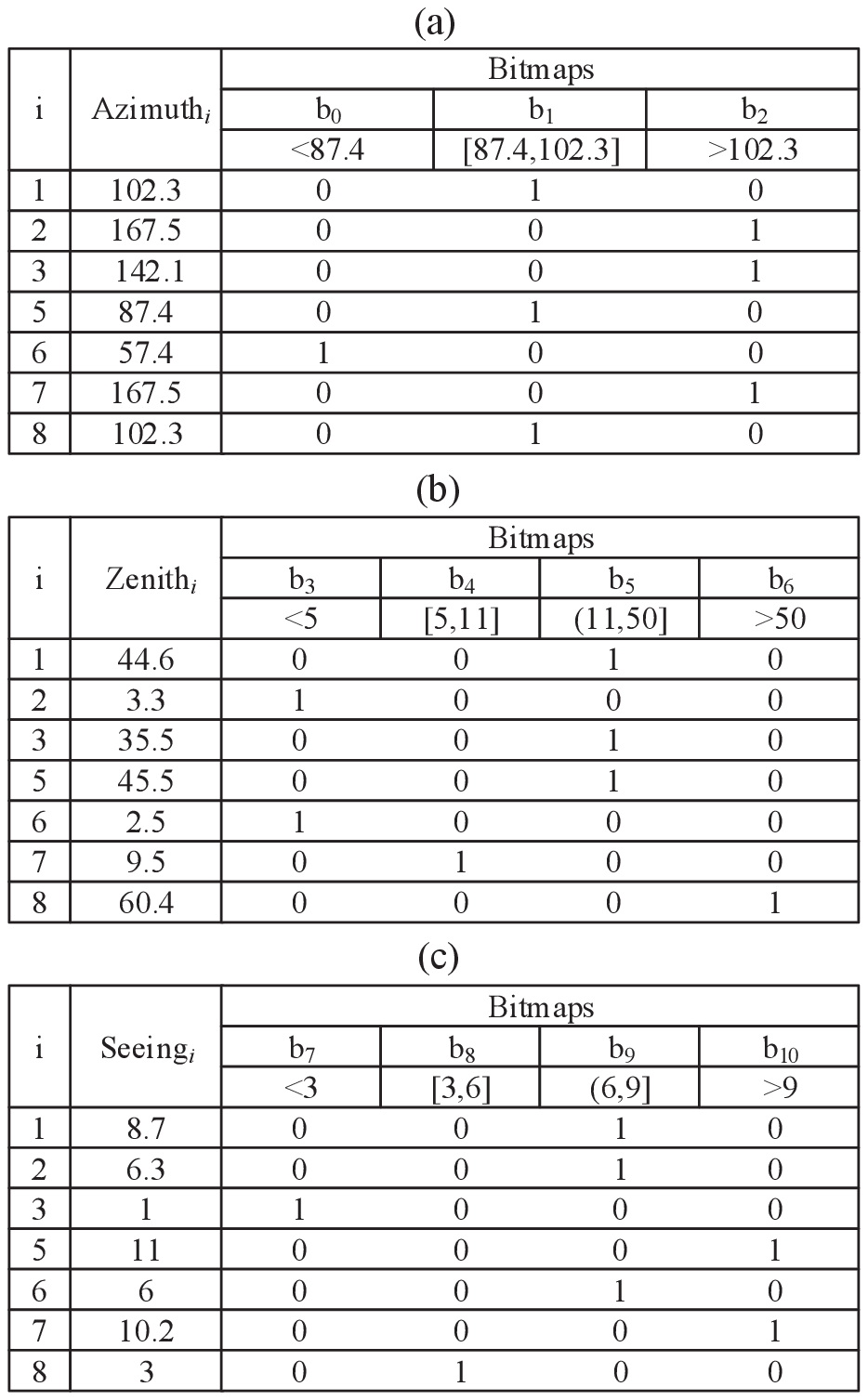}
\caption{Three simple bitmap indices for three attributes such as azimuth, zenith and seeing. The scales used for each attribute are just examples for explaining the principle of the bitmap index.\label{fig-wahexamp}}
\end{figure}

A bitmap index contains the same information as a B-tree. The difference is that the bitmap index technique replaces the row identifiers associated with each key value with a bitmap which can be operated efficiently by computer systems. For example, if we create a bitmap database with three fields, such as azimuth, zenith and seeing, the bitmap index technique would generate the index structure showed in Figure~\ref{fig-wahexamp}. For an attribute with ``azimuth'' values, the basic bitmap index generates azimuth bitmaps with 8 bits each, where 8 is the number of records (rows) in the data set. Each bit in the bitmap is set to ``1'' if the attribute in the record is of a specific value; otherwise the bit is set to ``0''. After the data indexing, there are three distinct categories and a 8-bit sequence is produced (see Figure~\ref{fig-wahexamp} (a)).

Using the bit sequence, it is easy to answer a range of queries. For example, query ``azimuth $<$= 102.3 and zenith $<$= 50 and seeing $>$= 10.0'' can be translated to the logical operations shown in Equation \ref{eq:Query},

\begin{equation}
\label{eq:Query}
\begin{split}
R=&(b_0\ or \ b_1)\\
    &and\  (b_3\ or\ b_4\ or\ b_5)\\
    &and\ b_{10}
\end{split}
\end{equation}

where \emph{R} is the bit sequence storing the result. If the \emph{i}th bit of \emph{R} is 1, then the \emph{i}th row of the record satisfies the query. It is easy to see that the bitmap index technique converts a conditional judgment from the search of field values to an operation on the bit sequence. This increases the query performance because the operations on bitmaps are well supported by computer hardware and the bitmaps can be combined easily and efficiently.

Obviously, the query arguments are typically arbitrary values that might not exactly coincide the range of indices of the bitmap during the actual data retrieval. Concerning this issue, many follow-up studies discussed the techniques of bitmap encoding, compression and binning, and further presented the corresponding solutions such as multi-component~\citep{neil1997improved} and multi-level bitmap encoding methods~\citep{wu2010multi}. Fastbit is capable of dividing the index range using either a linear scale and a log scale mode automatically. In general, the default Fastbit arguments  are enough to assure optimal database performance.

\subsection{EXPERIMENTS}
\label{test}
While the bitmap indexing can in theory be used for astronomical data archiving, few studies have presented implementations of bitmap indexing in astronomical contexts. Therefore, we have designed a series of tests to verify whether the bitmap index technique can meet the requirements of NVST. To test the performance of NoSQL database, we create a relational database by using MySQL with the same fields listed in Table \ref{tbl:queryfield} so as to prepare a comparison platform between the Fastbit database and relational database.
\begin{table}[t]
\begin{center}
\centering
\caption{Fields for Querying \label{tbl:queryfield}}
\doublerulesep2.0pt
\renewcommand\arraystretch{1.5}
\begin{tabular}{ll}
\toprule
Fields       &    Description  \\
\midrule
obs\_obj  &active region (AR) \\
obs\_date   & observation date\\
obs\_obj\_n  & coordinate north of AR\\
obs\_obj\_e & coordinate east of AR \\
obs\_org$^{\rm *}$ & observation station \\
azimuth  & azimuth of the telescope\\
zenith  & zenith of the telescope\\
number & instrument number\\
seeing & seeing\\
wind\_speed & wind speed\\
obs\_band  & observation band (e.g. TiO)\\
off\_band  & offset band (e.g. Center)\\
exp\_time  & exposure time of camera\\
temperature  & environment temperature \\
weather & weather\\
target & target (e.g. sunspot)\\
obs\_oper$^{\rm *}$ & observers\\
note$^{\rm *}$  & observation note \\
fits\_path$^{\rm *}$  & a reference to a FITS file\\
\bottomrule
\end{tabular}
\end{center}
\tabnote{$^{\rm *}$ Excluding from indexing. }
\end{table}
We simulate a total of 40 million test data files which are a monotonically increasing sequence with time to fit the real observational data as realisticly as possible. All data are imported to the Fastbit database and MySQL respectively according to the test requirements. In each test, the data sets for both databases are identical to guarantee the accuracy of the experiment.

The experimental server configuration specifications are an Intel (R) Core (TM) two Duo CPUs E7200 @ 2.53GHz with 4G RAM. The I/O subsystem is a disk with SATA2.0 7200 RPM.

\subsubsection{Index Creation Performance}
\begin{figure}[!htb]
\centering
\includegraphics[width=80mm]{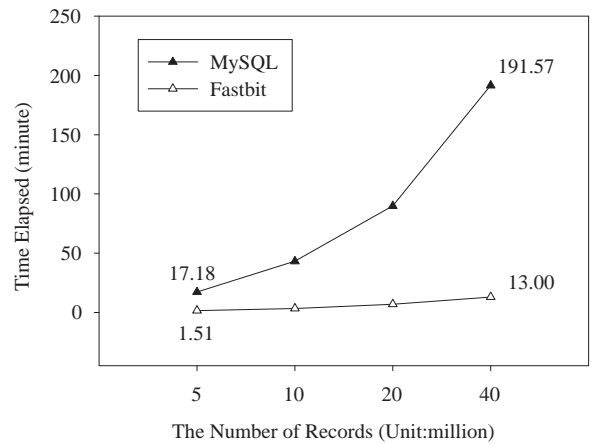}
\caption{The diagram of index creation. Time elapsed for index creation as a function of the number of records (in units of millions) for the MySQL (filled  triangles) and the Fastbit (open triangles) databases.\label{fig-indexcreat}}
\end{figure}

The first test is the index creation performance under the two databases. After importing the data, it is necessary to build database indexes to improve the query performance. For each database, we create indexes by subdividing the full range of the index in four parts. We repeat this process six times, with different subdivisions. The average index creation times are calculated and plotted in Figure~\ref{fig-indexcreat}. The time elapsed in creating the initial index for both databases is minute-level. Even on small data sets, MySQL takes about 17 minutes to build the initial index. However, Fastbit database only takes 1.51 minutes. The difference of index creation between the two kinds of database is about 10 times. With the increase of the data records, the difference will widen further.

In the case of a database crash, it would take a long time (more than 3 hours) to recover a relational database with 40 million records. On the contrary, the Fastbit database reflects a faster indexing speed. The index creation time of 13 minutes is acceptable for most database administrators.

\subsubsection{Data Insert Performance}
\begin{figure}[!htb]
\centering
\includegraphics[width=80mm]{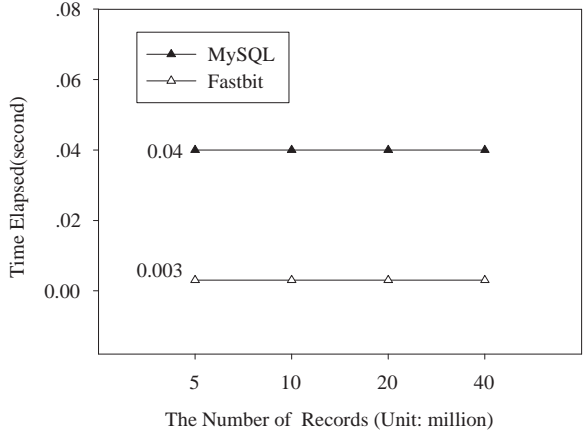}
\caption{The diagram of inserting a record. Time elapsed for inserting a record as a function of the number of records (in units of millions) for the MySQL (filled  triangles) and the Fastbit (open triangles) databases.\label{fig-insert}}
\end{figure}

It is very important to test the data insert performance because NVST produces massive amounts of data per day. The tests were conducted on the premise that each database has an index. The result shown in Figure~\ref{fig-insert} proves that the data insert is not an important issue in the DAS. Whether 0.04 seconds of MySQL or 0.003 seconds of Fastbit database, the data insert performance is far beyond the requirement of the NVST-DAS.

\subsubsection{Query Performance}
We conduct two kinds of query performance tests. One is a field query and the other is a multi-field combined query. Both query tests are similar to what astronomers would typically submit in their daily research work. To guarantee the test accuracy, we also repeat six times and record the query response time. The final result is an average value of the test records.

\begin{figure}[!htb]
\centering
\includegraphics[width=80mm]{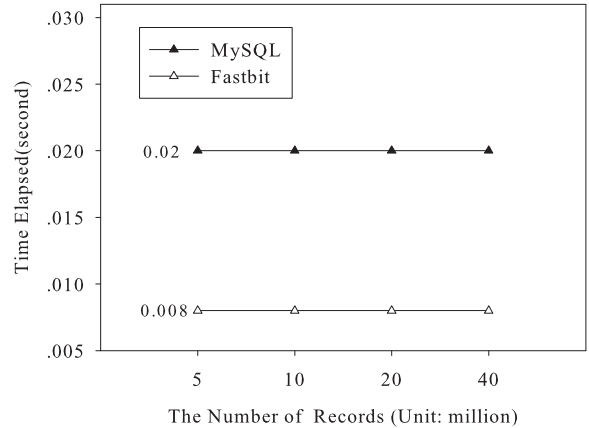}
\caption{The diagram of single field query. Time elapsed for single field query as a function of the number of records (in units of millions) for the MySQL (filled  triangles) and the Fastbit (open triangles) databases.\label{fig-expqryhits}}
\end{figure}

{\em1. Single field query.} Single field query is a common query requirement in astronomical data retrieval. For example, astronomers may submit a query condition with a range of observational time. This kind of query is only concerned with one field in the database. In the test, we submit a simple query criteria to query the data at a certain time from the two databases respectively. The test results are shown in Figure~\ref{fig-expqryhits}.

Obviously, both databases have extremely high performance in single field queries. Independent of the data scale, the query response time of Fastbit database remains at 0.008 second.

{\em2. Multi-field query.} Astronomers typically submit multiple parameters to query data from a database with combined criteria. We construct a SQL query statement, listed in Table~\ref{tbl:statements}, to simulate a routine data query for searching high quality observational data from the NVST. The average seeing (Fried parameter, r0) of NVST obtained in the period from 1998 to 2000 was about 10 cm \citep{Liu20111meter,liu2014new}. Therefore, it is very easy to query high quality data with three arguments, i.e., the azimuth (an azimuth angle from $180^{\circ}$ to $300^{\circ}$ means the Sun is on the Fuxian Lake), the zenith ($<50^{\circ}$) and the seeing ($>$ 12 cm) in observation.

\begin{table}[!htb]
\centering
\caption{SQL Query Statement. \label{tbl:statements}}
\renewcommand\arraystretch{1.5}
\begin{tabular}{|l|}
\hline
select obs\_obj, obs\_org, obs\_obj\_n, obs\_obj\_e,\\
obs\_date, obs\_oper, azimuth, zenith,\\
number, wind\_speed, obs\_band,\\
off\_band, seeing, exp\_time, temperature,\\
 weather, target, fits\_path, note \\
from tsolar t \\
where t.exp\_time/1000 $<$ 30.5 \\
and t.temperature $>$ 25.0   \\
and t.off\_band = 'Center'  \\
and t.obs\_band = 'TiO' \\
and target = 'Sunspot' \\
and t.zenith $<$ 50 \\
and t.azimuth $>$ 180 and t.azimuth $<$ 300 \\
and t.seeing $>=$ 12.0\\
\hline
\end{tabular}
\end{table}

\begin{figure}[!htb]
\centering
\includegraphics[width=80mm]{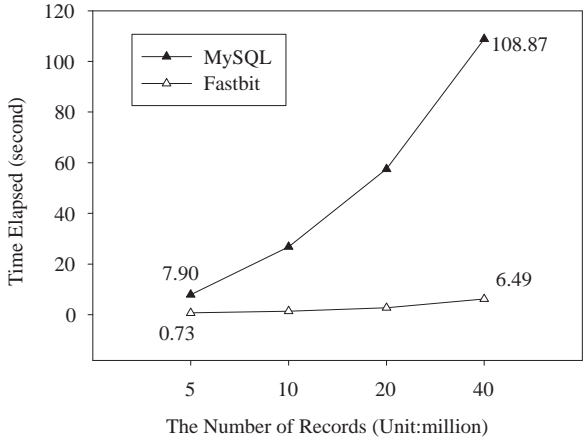}
\caption{The diagram of multi-field query. Time elapsed for multi-field query as a function of the number of records (in units of millions) for the MySQL (filled  triangles) and the Fastbit (open triangles) databases.\label{fig-experquery}}
\end{figure}

Figure~\ref{fig-experquery} shows that the test results of the multi-field combined query performance are significantly different from the results of the single field query. In the case of small data sets (e.g., 5 million records), the query response times of the two databases are basically the same. As the data amount increases, the multi-field query response time of the relational database grows exponentially. However, the corresponding query response time for the Fastbit database does not change significantly as a function of the number of records. Even if the amount of records reaches 40 millions, the query response time is still less than 10 seconds.

\subsubsection{Storage Space Comparison}
\begin{figure}[!htb]
\centering
\includegraphics[width=80mm]{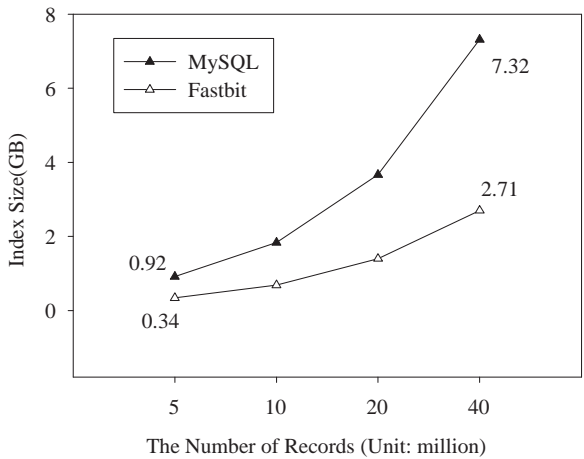}
\caption{The diagram of the disk space allocation. Index size for disk space allocation as a function of the number of records (in units of millions) for the MySQL (filled  triangles) and the Fastbit (open triangles) databases.\label{fig-indesize}}
\end{figure}

The storage space is not a critical issue in the construction of the DAS because the price of the storage equipments is becoming cheaper and cheaper. However, for a data archiving system, it is necessary to consider reducing the overhead of allocating disk space. In this test, we analyze the storage space allocation of each database. Figure~\ref{fig-indesize} shows the allocation values for different data scales. Surprisingly the size of the Fastbit database is smaller than that of the relational database. For the same amount of data, the Fastbit database can save about half a million more records.

\subsubsection{Summary}
Based on the  presented performance experiments, we obtain some suggestions about the decision of the database technology.

1) MySQL has good performance on small data sets, but the query response time grows exponentially with increasing records. It would take more than 100 seconds to query data from a database with more than 40 million records. Obviously, the query performance is inefficient to support subsequent data processing issues. Although a large database can be split into several small databases so as to improve the query performance, the database administrator would face big challenges of database maintenance and upgrade.

2) Fastbit NoSQL database is proved to be a potent database technology for massive data queries. It uses the compressed word-aligned bitmap index scheme to achieve high-performance data for indexing.  For all the tests conducted, Fastbit database shows distinct advantages in index creation, query performance and storage space allocation. These results prove that Fastbit performs more efficiently than MySQL under massive data loads. Although there are only few successful cases that can be referenced, it is worth attempting to develop a high performance data archiving system utilizing the Fastbit database technology.

\section{THE NVST DATA ARCHIVING SYSTEM}

\subsection{System Architecture}

\begin{figure}[!htb]
\centering
\includegraphics[width=80mm]{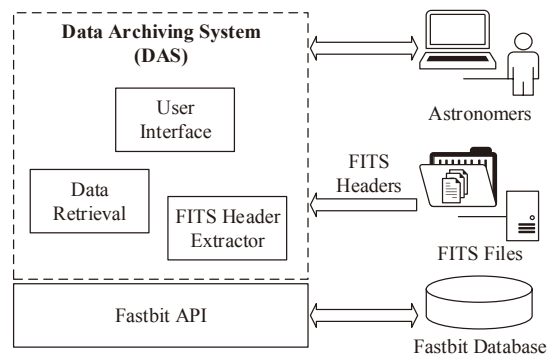}
\caption{The diagram of the DAS architecture.  A data extraction sub-system extracts the observational parameters and related information (meta data) from FITS files. A data retrieval sub-system provides web-based Fastbit database searching. Astronomers submit requests to DAS through the user interface. The DAS uses the API of Fastbit to perform queries and delivers satisfied records to astronomers.\label{fig-archi}}
\end{figure}

As an information management system, the DAS of the NVST fully depends on the FastBit compressed bitmap index technique instead of a traditional relational database. Figure~\ref{fig-archi} shows an architecture diagram that describes how the DAS processes data. The DAS mainly consists of two sub-systems. One is a data extraction sub-system which extracts the observational parameters and related information (meta data) from FITS files. The other one is the data retrieval sub-system that provides web-based Fastbit database searching.  DAS is written in C++ language and provides several software interfaces that can be invoked by any other programming languages such as Python and Java for future function expansion.

\subsection{FITS Header Extractor}

Unlike relational databases, while FastBit also treats data as tables with rows and columns, Fastbit makes an index for each column instead of making an index for each row in a relational database. According to the requirements of astronomers, we organize all columns (same as the field in a relational database) that are possibly queried in DAS and listed in Table \ref{tbl:queryfield}.

\begin{figure}[!htb]
\centering
\includegraphics[width=80mm]{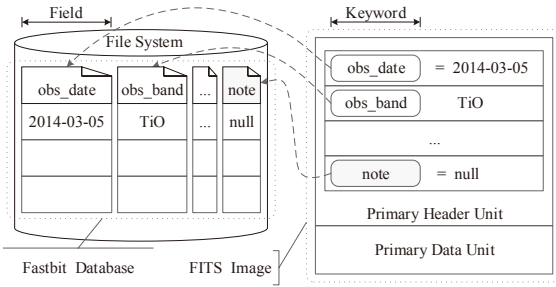}
\caption{Extracting FITS header from FITS file into DAS Fastbit NoSQL database.\label{fig-extract}}
\end{figure}

To obtain the data from FITS file, the FITS Header Extractor (FHE) module is implemented. FHE traverses the specific directories recursively and searches all FITS files. With the support of the CFITSIO library\footnote{http://heasarc.gsfc.nasa.gov/fitsio/fitsio.html}, FHE is used to open the FITS file, extract the information (meta data) from the header data unit (HDU) of the FITS file and finally insert the meta data into the Fastbit database. Figure~\ref{fig-extract} shows a diagram of the extraction of the meta data.

The Fastbit database is stored in a directory of the file system, with each column stored as a separate file in raw binary form. The name of each column file is the name of the column. There must be a meta data file named -part.txt in the directory for a data partition. This file contains information such as the name of the partition, the number of rows in the partition, the number of columns and column names.

\subsection{Data Retrieval}
\begin{figure}[!htb]
\centering
\includegraphics[width=80mm]{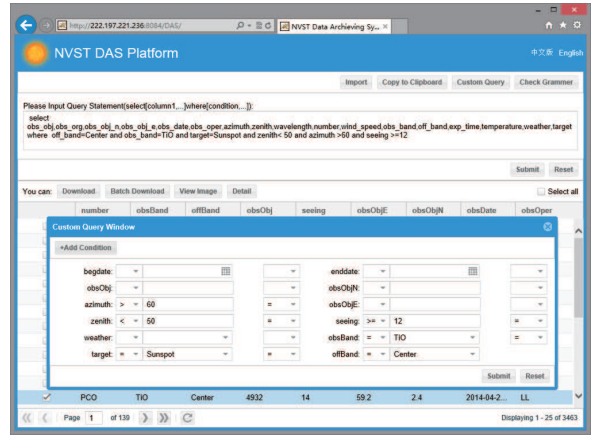}
\caption{The DAS provides a friendly user interface for astronomers to construct complicated query statements by selecting the query fields in a ``Custom Query" window. \label{fig-dasportal}}
\end{figure}

The DAS provides a web interface (see Figure~\ref{fig-dasportal}) for astronomers. The users can input the query statements with SQL-like style. The DAS provides a friendly user interface for astronomers to construct complicated query statements by selecting the query fields in a ``Custom Query" window, and the statements can be automatically built. Some useful functions, such as importing a text file of statements, checking the grammar of query statements and so on, can be executed by the users to simplify their interactions with the query system. After a query, the astronomers can download one or a batch of files according to their needs and priorities.

\section{CONCLUSION}

The construction of the DAS is one of the most significant issues in the NVST data processing. In this study, we implement a Fastbit NoSQL database instead of the traditional relational database and provide the full functions to support web-based high performance query for the astronomers. A series of experiments is the designed to test the performance differences between two kinds of databases. The test results indicate that Fastbit indexing has outstanding performance for astronomical data, especially for very large data sets.

It also should be pointed out that the development of the DAS with Fastbit database technology is more difficult than that with relational database. The Fastbit library provides only a collection of functions or software interfaces which can be called. There are only a few tools to help the developers to check and verify the final results. One of the most difficult things to develop is the concurrency of the Fastbit database. The current version of the Fastbit database cannot support reading and writing records in one database concurrently. This means that the observational data cannot be inserted into the database in real time. Although this is not a big problem for a data archiving system, it may complicate the work of the database administrator.

Overall, we can conclude that the NVST DAS is a successful case using the Fastbit NoSQL database. The outstanding performance of the Fastbit technology provides more varied opportunities for subsequent massive astronomical data processing of the NVST. Our study would be a valuable reference for other modern telescopes and would further contribute to the community of massive astronomical data processing.

\acknowledgments
We acknowledge the support from the National Natural Science Foundation of China (No. U1231205, 11263004, 11203011, 11163004 and 11103005).



\end{document}